\documentclass[aip,apl,preprint]{revtex4-1}
\usepackage{graphicx}
\usepackage{natbib}

\begin{document}	
 
	\title{Feedback-Controlled Electromigration for the Fabrication of Point Contacts}
	\date{\today}
	
	\author{J.M. Campbell}
	\author{R.G. Knobel}
	\email[Electronic mail:]{knobel@physics.queensu.ca}
	\affiliation{Department of Physics, Engineering Physics and Astronomy, Queen's University, Kingston, Ontario, Canada}

	\begin{abstract}
	Lithographically fabricated point contacts serve as important examples of mesoscopic conductors, as electrodes for molecular electronics, and as ultra-sensitive transducers for mechanical motion. We have developed a reproducible technique for fabricating metallic point contacts though electromigration.  We employ fast analog feedback in a four-wire configuration in combination with slower computer controlled feedback to avoid catastrophic instability.  This hybrid system allows electromigration to proceed while dissipating approximately constant power in the wire.  We are able to control the final resistance of the point contact precisely below 5~k$\Omega$ and to within a factor of three when the target resistance approaches 12~k$\Omega$ where only a single conducting channel remains.
	\end{abstract}

	\maketitle

Thin wires display interesting phenomena as they are narrowed to the breaking point, and can form useful transducers.  Such wires have been used as the basis for much of molecular electronics~\cite{park1999,strachan2005,fischbein2006,johnston2007} and can exhibit mesoscopic phenomena including the Kondo effect~\cite{houck2005}.  As a wire is narrowed at a constriction, its resistance increases until only a few conducting channels remain, forming a point contact or a quantum wire. In this regime the point contact is described by Landauer's scattering formalism in which the conductance is a multiple of $G_0 = 2e^2 / h $ ~\cite{agrait2003}.  As the wire is further narrowed, the resistance increases in discrete steps until it is broken and a nanometer-scale gap or tunneling gap is created.   The tunneling across such a gap forms the heart of the scanning tunneling microscope~\cite{binnig1982_stm}.  On-chip point contacts have the potential to be transducers for high-bandwidth, ultra-sensitive mechanical measurements~\cite{flowers2007}.

Point contacts can be fabricated through scanning probe methods~\cite{gimzewski1987,xu2003}, mechanically-controlled break junctions (MCBJ)~\cite{moreland1985,zhou1995}, and electromigration (EM)~\cite{park1999}.  While scanning probe methods and MCBJ employ mechanical apparatus, EM requires only electrical contact with the device making it an attractive technique for the fabrication of point contacts, if only it were more reproducible.  EM involves applying a voltage $V$ across a lithographically patterned wire that narrows to a neck.  The voltage is increased until the neck, where current density $j$ is highest, reaches a critical temperature $T^*$ through Joule heating.  At that point, the metallic atoms become mobile and, if the current density is sufficiently high, momentum transfer from the conduction electrons to atoms results in a net drift of atoms which narrows the wire~\cite{park1999}.  Reproducible fabrication of point contacts through EM with a simple voltage ramp, however, is difficult due to the rapidly changing temperature and current density in the wire~\cite{park1999}.  Thermal runaway tends to result in catastrophic breaking of the wire leaving a large gap instead of a point contact~\cite{esen2005,taych2007}. 

\begin{figure}
		\centering
			\includegraphics[width=85mm]{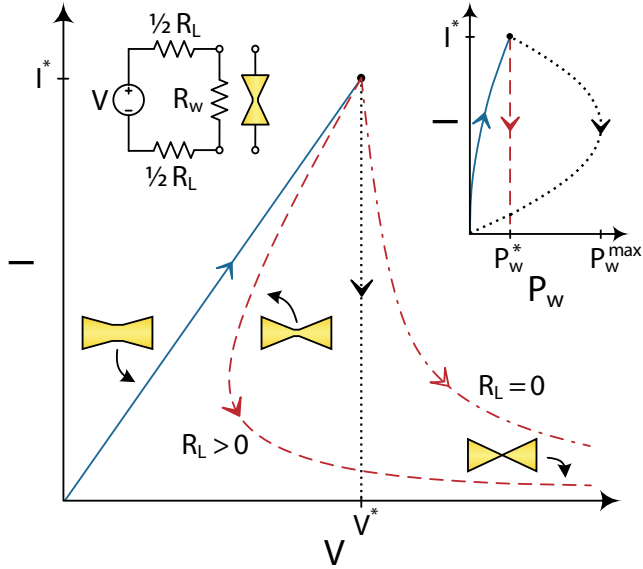}
		\caption [] {Current-voltage characteristic during EM for a wire with resistance $R_w$.  The voltage $V$ is increased until the wire reaches a critical temperature $T^*$.  If the lead resistance $R_L > 0$, the $IV$ relation is double-valued and the process is unstable.  The instability is removed if $R_L = 0$.  The inset shows power dissipated in the wire $P_w$ during the process.  As $V$ is initially increased, the power increases until a critical power $P_w^*$ is attained.  When EM is controlled, $P_w$ is constant (red dashed line).  If EM is not controlled and the applied voltage is held constant, the power dissipated in the wire exceeds the critical power $P_w^*$ (black dotted line). (Parts of figure adapted from ref.~\cite{wu2007}.)}
		\label{fig:IVcharacEM}
	\end{figure}

Uncontrolled EM is induced by the lead resistance between the voltage source and the point contact junction~\cite{taych2007}.  Figure~\ref{fig:IVcharacEM} shows the current--voltage relationship during EM when there is non-zero series resistance.  In the initial phase of the process, the voltage ramp results in Joule heating but there are no permanent morphological changes to the junction.  Once sufficient power $P_w^*$ is dissipated in the wire, a critical temperature $T^*$ is reached and EM begins.  Atoms diffuse or are pushed away from the neck, causing it to narrow and increase in resistance.  Controlled EM takes place when the power dissipated in the wire $P_w = V I - R_L I^2$ is approximately constant~\cite{houck2005,strachan2005} or decreasing~\cite{wu2007,Stoffler2012}.  For non-zero lead resistance $R_L$, the constant power condition results in an unstable, double-valued $IV$ relation.  As shown in the inset of Figure~\ref{fig:IVcharacEM}, even if the voltage is held constant once $P_w^*$ is reached, $P_w$ increases to approximately $P_w^* R_L / 4 R_w^0$ where $R_w^0$ is the wire resistance before EM begins.  The rapid increase in temperature and current density usually results in explosive breaking of the wire~\cite{taych2007}.  

There are two previously used techniques to improve control of the EM process.  The first employs computer controlled feedback to monitor the wire resistance and adjust the applied voltage to keep the dissipated power approximately constant~\cite{strachan2005}.  Software feedback is slow, however, and such feedback isn't enough to ensure stability if the series resistance is large.  It might not be possible to minimize lead resistance in some cases: low-temperature cryostats, for example, often have resistive wiring to reduce heat transfer.  An alternative technique is to remove series resistance effects by lithographically patterning four-terminal devices and adding analog feedback to control the voltage directly across the wire~\cite{wu2007}.  In this way, the $IV$ relation is no longer double-valued and the instability in the EM process is reduced.  This method does place restrictions on the circuit design, however.  If the point contact were to be used to measure the displacement of a mechanical oscillator, for example, it might not be possible to pattern four wires directly down to wire.  Furthermore, radio frequency (RF) measurement techniques (for example, ref.~\cite{Puebla-Hellmann2012a}) do not allow for easy four-wire patterning directly to the wire.

In order to increase our control over the EM process in a way that is compatible with point contact displacement detectors, we have developed a hybrid feedback scheme consisting of both hardware and software components.  We combine computer-controlled feedback with fast analog feedback in a four-wire configuration beginning at our lithographically-patterned contact pads.  This removes the effects of the relatively large cryostat wiring resistance without placing restrictions on device design.  This hybrid technique allows us to reproducibly fabricate point contacts with some control over final resistance even when the series resistance is significant.  Finite element modeling was used to study the relationship between wire geometry, current density and temperature during the EM process.

\begin{figure}
		\centering
			\includegraphics{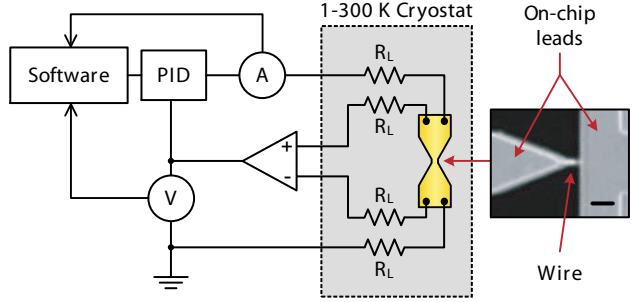}
		\caption {Schematic of the hybrid feedback system in a four-wire configuration.  The PID controller (SRS-SIM960) monitors the voltage across the point contact junction as measured with a differential amplifier (Signal Recovery 5113 pre-amp) and adjusts its output to keep this voltage at a given setpoint.  The voltage across the junction as well as the current though it are monitored (Agilent 34970, HP 34401) by a computer which then controls the PID setpoint according to the feedback algorithm.  The scale bar is 500~nm.}
		\label{fig:hybridEMcircuit}
	\end{figure}

Gold nanoscale wires and larger on-chip leads for EM were fabricated on a 50~nm thick silicon nitride layer on a silicon substrate through electron beam lithography, double-angle evaporation and a lift-off process.  The on-chip leads were composed of 80~nm of gold with a 5~nm chromium sticking layer.  The wires were $50\times 200$~nm ($w\times l$) of 30~nm  thick gold.   Double-angle evaporation allowed the gold wire, where EM is to occur, to be deposited directly on the substrate without the chromium sticking layer and also to be thinner than the on-chip leads.  A typical device is shown in Figure~\ref{fig:hybridEMcircuit}.

The hardware portion of the hybrid feedback scheme involves a four-wire setup similar to that by Wu \textit{et al}.~\cite{wu2007} which removes the effect of the relatively large lead resistance of our 1-300~K cryostat (see Figure~\ref{fig:hybridEMcircuit}).  The voltage source is a PID controller which, when combined with the four-wire measurement, allows a target voltage to be set across the device itself.  The PID controller has a bandwidth of 100 kHz, which allows for small corrections to the device voltage at a rate much faster than what is provided by the software feedback. 

While the hardware feedback increases the stability dramatically, a slower software feedback allows a controlled approach to the desired resistance.  The software feedback scheme is similar to that developed by Strachan \textit{et al}.~\cite{strachan2005} in which the voltage applied across the device is computer controlled and depends on the resistance of the wire.  We use the breaking rate $R^{-1}dR/dt$ as an additional control variable~\cite{houck2005}.  The algorithm is as follows: The initial resistance of the wire is measured and the voltage is increased at 4 mV/s. When the resistance increases by a percentage to $R_p$ or when $R^{-1}dR/dt$ exceeds a threshold, the voltage is decreased by 10~mV to stop EM. This procedure is repeated until a target resistance $R_t$ is reached.

The hybrid feedback scheme was tested with several devices in both ambient conditions and at 77~K.  The devices, similar to that shown in Figure~\ref{fig:hybridEMcircuit}, had initial resistances between 35 and 55~$\Omega$.  The target resistance was between 1 and 12~k$\Omega$ which, according to the Landauer formula, should leave just a few conduction channels between the on-chip leads.  In this resistance regime, control of the EM process was found to be most difficult. 

\begin{figure}
		\centering
			\includegraphics{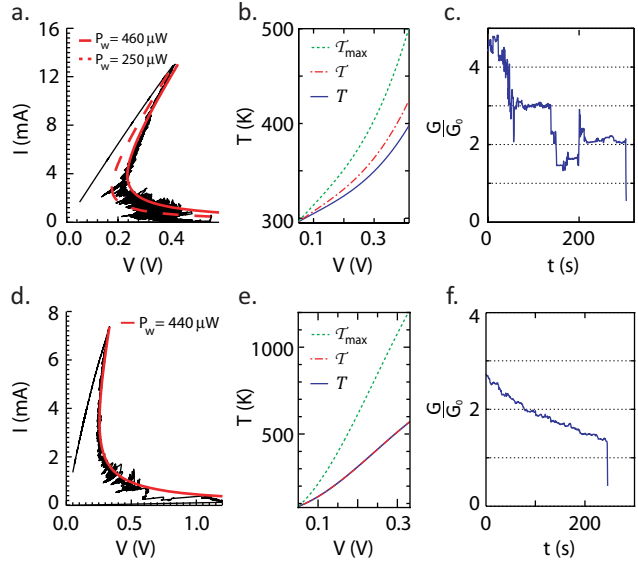}
		\caption {EM results for (a-c) a 108~nm wide wire in ambient conditions and (e-f) a 77~nm wide wire at 77~K. (a,d) IV characteristic during EM.  The solid and dashed curves show constant power dissipated in the wire.  (b,e) The average wire temperature $T$ calculated from the temperature coefficient of resistance $\alpha$ and the average temperature $\mathcal{T}$ and maximum temperature $\mathcal{T}_{max}$ of the wire determined from the temperature--resistance relationship of a finite element simulation.  (c,f) Wire conductance at the end of the EM process.}
		\label{fig:EMresults}
	\end{figure}

The relationship between the measured current and sourced voltage during the EM process for devices at both room temperature and 77~K is shown in Figure~\ref{fig:EMresults}a,d.  For both cases, the resistance initially changed only slightly as the increase in voltage heated the wire.  Then, when the wire reached a critical temperature $T^*$, EM began.  In ambient conditions, this occurred at $0.56 \pm 0.08$~V while at 77~K, the critical point was slightly lower at $0.36 \pm 0.08$~V.  This difference was expected as the cryogenic temperatures reduced the on-chip lead resistance $R_L$.  As the resistance of the wire increased due to EM, the feedback kept the power dissipated in the wire roughly constant (Fig.~\ref{fig:EMresults}b) or decreasing (Fig.~\ref{fig:EMresults}a).  At 77~K, $R_L$ is significantly reduced, removing much of the reentrant nature of the $IV$ curve, and the instability of the EM process.  Near the end of the EM process, steps in the conductance in rough multiples of the conductance quantum $G_0 = 2e^2/h$ appeared, indicating that only a few conduction channels remained (see Fig.~\ref{fig:EMresults}c,f).  In addition to steps, one device also displayed a smooth decrease in conductance to $G_0$ (see Fig.~\ref{fig:EMresults}f).  This is likely due to impurities in the junction that result in a transmission probability less than unity.
	
\begin{figure}
		\centering
			\includegraphics{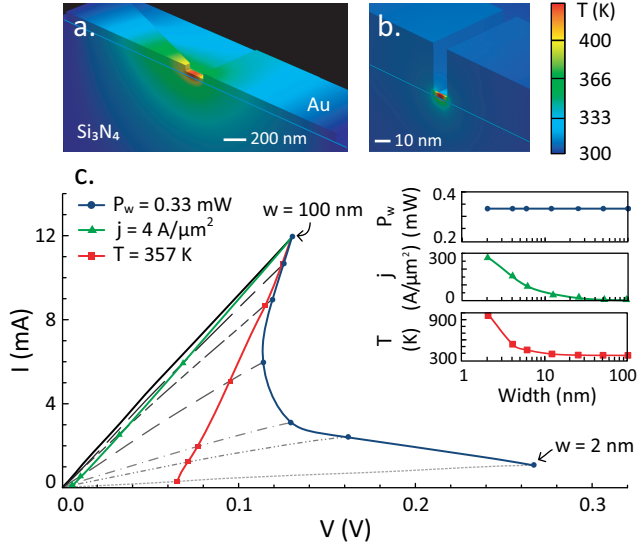}
		\caption [FEM IV characteristic] {Temperature contours for a finite element model of a) a 100~nm wide wire and b) a 2~nm notch in the wire.  Half of the model is shown as the forward facing plane is a symmetry plane.  c) $IV$ characteristics for a finite element simulation of a $30\times 100\times 200$~nm ($t\times w\times l$) Au wire in Coventor~\cite{coventor}.  The cross section of the center of the Au wire was decreased from $100 \times 30$~nm to $2 \times 2$~nm over a 10~nm length.  Curves of constant power $P_w$ in the Au wire, maximum temperature $T$ and maximum current density $j$ were calculated beginning where $j=4$~A/$\mu$m$^2$.  The inset shows the temperature and current density along the curve of constant dissipated power.}
		\label{fig:coventorSimIV}
	\end{figure}

All devices displayed slight curvature of the $IV$ relation during the initial phase of the process indicating an increase in resistance.  Since resistance changes are reversible along this branch of the $IV$ characteristic, EM has not yet begun and there are no morphological changes to the device. Therefore, we can infer that the increase in resistance is due to rising temperature in the wire.  As shown in Figure~\ref{fig:coventorSimIV}a, finite element simulations indicate that the increase in temperature is mostly confined to the thin Au wire; there is minimal heating of the on-chip leads.  For a metal, an increase in temperature $\Delta T$ will result in a fractional change in resistance of $\Delta R/R_0 = \alpha \Delta T$ where $R_0$ is the initial resistance and $\alpha$ is the temperature coefficient of resistance.  For our evaporated Au wires, we measure $\alpha = 0.002$ allowing us to calculate a rise in average wire temperature $T$ to $420 \pm 20$~K and $540 \pm 70$~K for EM at room temperature and 77~K, respectively.  These estimates are similar to those reported elsewhere~\cite{strachan2008,taych2007,trouwborst2006}.  We also estimated the critical current density at the onset of EM using the width of the Au wires as measured with a scanning electron microscope.  The average critical current density was $4.1 \pm 0.7$~A/$\mu$m$^2$, which confirms other results~\cite{trouwborst2006}.  There was no statistically significant difference between the critical current density in room temperature and 77~K devices.

Once EM began and the wire morphology started to evolve, the temperature and current density could no longer be estimated as above.  Previous studies showed that temperatures determined from resistance changes are no longer accurate when the effective wire length is comparable to the inelastic scattering length, estimated as 20~nm for similar Au films~\cite{wu2007}.  We created a finite element model of the device to study the relationship between junction geometry, current density and temperature during EM.  The model consisted of a $30\times 100\times 200$~nm ($t\times w\times l$) Au wire connecting two approximately $1.5 \times 1$~$\mu$m Au leads on a silicon nitride substrate.  We input the measured electrical conductivity of similar Au films; thermal conductivity was determined from electrical conductivity through the Wiedemann-–Franz law.  The initial phase of the process, up to the onset of EM, was simulated by increasing the voltage across the electrodes.  The relationship between the subsequent increase in resistance and temperature was used to estimate the average wire temperature $\mathcal{T}$ and maximum wire temperature $\mathcal{T}_{max}$ of experimental devices (see Fig.~\ref{fig:EMresults}b,e).  $\mathcal{T}$ was similar to the temperature $T$ calculated from $\alpha$. $\mathcal{T}_{max}$, which was about 1.5~times greater than $\mathcal{T}$, is a better estimate of the temperature at the wire location where EM occurs.  From our simulations, $\mathcal{T}_{max} = 590 \pm 50$~K at the onset of EM.  Therefore, an electromigrated point contact may not always be suitable for use as electrodes for molecular electronics as some molecules of interest could be damaged at this temperature~\cite{heersche2006}.      

To simulate EM, the center of the Au wire was then narrowed and thinned over a 10~nm length from a cross-section of $100 \times 30$~nm to $2 \times 2$~nm.  Curves of constant power $P_w$ in the Au wire, maximum temperature and maximum current density $j$ were calculated beginning where $j=4$~A/$\mu$m$^2$.  The combination of the IV curve for the 100~nm wide wire and the curve of constant power provide an IV characteristic for simulated EM.  Results, shown in Figure~\ref{fig:coventorSimIV}, indicate that the Au wire narrows rather quickly along the constant power curve reaching the nanometer range near the bottom of the curve.  It is at this point where the feedback system does not follow the constant power curve as closely and in several cases, the dissipated power decreases.  However, as the wire narrows, bulk models of electrical and thermal conduction become increasingly inaccurate and individual metallic grains and defects play a dominant role.   

\begin{figure}
		\centering
			\includegraphics{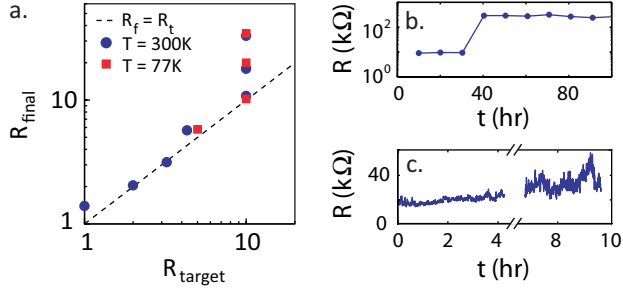}
		\caption [Controlling the final resistance] {Final device resistance for various target resistances for EM in ambient conditions as well as at 77~K.  The sourced voltage began at 50~mV and the feedback condition was $R^{-1}dR/dt < 0.025 s^{-1}$.  b-c) Evolution of wire resistance after completion of the EM process for devices at room temperature.}
		\label{fig:EMresultsSum}
	\end{figure}

Next we studied our ability to control the final resistance of a device through feedback-controlled EM.  We carried out several EM runs in ambient conditions as well as at 77~K.  Results are shown in Figure~\ref{fig:EMresultsSum}a.  For target resistances below 5~k$\Omega$, the final resistance was nearly equal to the target resistance.  When the device narrows near the point of a single atomic chain, it is more unstable and the resistance is typically greater than the target resistance but no greater than around a factor of three.  Stability of the electromigrated devices at room temperature is minutes to hours (see Fig.~\ref{fig:EMresultsSum}b-c), while devices at low temperature have longer stability. 

This hybrid technique for fabricating point contacts through electromigration allows us to reproducibly fabricate point contacts with some control over final resistance even when there is significant series resistance.  This system makes it possible to efficiently create point contacts through electromigration for fundamental studies of atomic-size conductors~\cite{agrait2003,houck2005} or applications such as displacement transducers.  The hybrid circuit allows for high device yield without an on-chip four-wire configuration or without minimizing series resistance.  This will facilitate experiments involving more complicated circuits such as microwave impedance matching~\cite{Puebla-Hellmann2012a} and investigating excess back-action in point contact displacement transducers~\cite{flowers2007,bennett2010}.

	\begin{acknowledgments}
	We thank Ben Lucht and Devon Stopps for work on early experiments.  This work was supported by Queen's University, Natural Sciences and Engineering Research Council of Canada, Canada Foundation for Innovation, Ontario Innovation Trust and Ontario Ministry of Training, Colleges, and Universities. 
	\end{acknowledgments}


%

\end{document}